\documentclass[doublecol,linenumbers]{epl2} 
\usepackage{amsfonts,amssymb,latexsym,amsmath,enumerate,amsthm}

\def\Im{{\rm Im}}
\def\Re{{\rm Re}}

\newcommand{\be}{\begin{equation}}
\newcommand{\ee}{\end{equation}}
\newcommand{\bea}{\begin{eqnarray}}
\newcommand{\eea}{\end{eqnarray}}

\revision

\title{Probing phase of a scattering amplitude beyond the plane-wave approximation}
\shorttitle{Probing phase of a scattering amplitude beyond the plane-wave approximation} 

\author{Dmitry Karlovets\inst{1}
}
\shortauthor{D. Karlovets}

\institute{                    
  \inst{1} Department of Physics, Tomsk State University, Lenina Ave. 36, 634050 Tomsk, Russia
}

\abstract{
Within a plane-wave approach, a number of scattering events in a collision is insensitive to a general phase of a transition amplitude, although this phase is extremely important for a number of problems, especially in hadronic physics. In reality the particles are better described as wave packets, and here we show that the observables grow dependent upon this phase if one lays aside the simplified plane-wave model. We discuss two methods for probing how the Coulomb- and hadronic phases change with a transferred momentum $t$, either by colliding two beams at a non-vanishing impact-parameter or by employing such novel states as the vortex particles carrying orbital angular momentum or the Airy beams. For electron-electron collision, the phase contribution to a cross section can reach the values higher than $10^{-4} -– 10^{-3}$ for well-focused beams with energies of hundreds of keV.}

\begin{document}

\maketitle

\section{Introduction} 

In a quantum theory of scattering with all states being unlocalized plane-waves, the cross section $d\sigma$ is known to be independent 
of the phase $\zeta_{fi}$ of a transition amplitude $T_{fi} = |T_{fi}|\exp\{i\zeta_{fi}\}$.
It is so as long as we neglect finite sizes of the wave packets and beams, their spreading during the collision, 
and suppose that the momentum uncertainties $\sigma$ are vanishing. But for several important exceptions \cite{Impact, t1, t2, Akhmedov_09, Akhmedov_10, Akhmedov_Found}, 
this approximation works very well. On the other hand, this phase turns out to be of high importance for hadronic physics, especially for $pp$ and $p\bar{p}$ collisions,
in which it is extracted from interference between the Coulomb- and the hadronic parts of the amplitude (see, for example, \cite{Y} and also \cite{Dremin} for a modern review).
The function $\rho = \Re T_{fi}/\Im T_{fi} = 1/\tan \zeta_{fi}$ is calculated within different models, including Regge approaches, 
and it is believed to depend in a complex way on the energy and on the transferred momentum. 
This phase has been recently extracted from measurements by the TOTEM collaboration at the LHC at $\sqrt{s} = 7$ TeV \cite{TOTEM},
and it has been found that, in agreement with the unitary considerations, the amplitude becomes mostly imaginary for the small scattering angles \cite{TOTEM_phase}. 
A further analysis of the data showed, however, that at large transferred momenta a real part of the amplitude can dominate \cite{Dremin_PRD}.

\begin{figure}
\centering
\includegraphics[width=8.80cm]{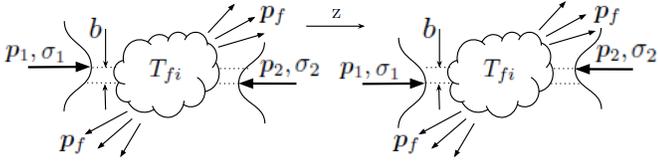}
\caption{For measuring the asymmetry, one compares outcomes of the two experiments with the swapped beams 
or, when the impact parameter ${\bm b}$ is vanishing, with the beams having opposite signs of the phases.
Alternatively, one can carry out just one experiment and measure the up-down asymmetry in the angular distributions.} 
\label{Fig}
\end{figure}

Here we show that the observables become sensitive to the phase $\zeta_{fi}$
if one lays aside the simple but unrealistic plane-wave approximation and treats all the particles 
as spatially- and temporarily localized wave packets.
Let in a general $2\rightarrow N_f$ scattering process with two identical incoming beams ($ee \rightarrow X, pp \rightarrow X$, etc.)
a ratio $\lambda_c/\sigma_b$ serve as a Lorentz-invariant small parameter, where $\lambda_c =\hbar/(mc)$ is a particle's Compton wave length and $\sigma_b$ is a beam's width.
Then the cross section, $d\sigma = dN/L$, represents a series in powers of $(\lambda_c/\sigma_b)^2$:
\begin{eqnarray}
& \displaystyle
d\sigma = d\sigma^{(pw)} + d\sigma^{(1)} + \mathcal O((\lambda_c/\sigma_b)^4),
\label{Eq0}
\end{eqnarray}
where $d\sigma^{(1)} = \mathcal O((\lambda_c/\sigma_b)^2)$, $dN, L$ are the number of events and the luminosity, respectively, 
$d\sigma^{(pw)} = (2\pi)^4 \delta^{(4)} (\langle p\rangle_1 + \langle p\rangle_2-p_f)\, |T_{fi}|^2 \,dn_f /\upsilon$ is the standard plane-wave cross section.
This first correction to the latter, $d\sigma^{(1)}$, vanishes in the plane-wave limit with $\sigma_b \rightarrow \infty$ 
and depends upon the phase $\zeta_{fi}$, 
on an impact parameter ${\bm b}$ between the beams' centers (a head-on collision is implied for simplicity), 
and on the phases $\varphi_1 ({\bm p}_1), \varphi_2 ({\bm p}_2)$ of the beams' wave functions 
$\psi_{1,2} ({\bm p}_{1,2})$. 
To be more precise, the cross section depends upon a combination
\begin{eqnarray}
\displaystyle
{\bm b}_{\varphi} - \left (\frac{\partial}{\partial {\bm p}_1} - \frac{\partial}{\partial {\bm p}_2}\right ) \zeta_{fi},\quad 
{\bm b}_{\varphi} = {\bm b} - \frac{\partial \varphi_1}{\partial {\bm p}_1} + \frac{\partial \varphi_2}{\partial {\bm p}_2}.\
\label{Eq0a}
\end{eqnarray}
As we demonstrate below, one can probe the phase $\zeta_{fi}$, or rather its derivative $\partial \zeta_{fi} (s, t)/\partial t$, 
with $s = (p_1 + p_2)^2,\, t = (p_1 - p_3)^2$, by comparing cross sections with a flipped sign of ${\bm b}_{\varphi}$.
This can be realized either by swapping the beams, i.e. ${\bm b} \rightarrow -{\bm b}$ (see Fig.\ref{Fig}), or by changing signs of the phases $\varphi_{1,2}$.


Whereas conventional Gaussian beams are needed in the first scenario, the second one requires more sophisticated quantum states.
The beams with the phases, i.e. $\psi ({\bm p}) \propto \exp\{i \varphi ({\bm p})\}$, are not plane waves, even approximately, 
and their wave functions in configuration space may turn out to be non-Gaussian. 
Depending on these phases, such states can represent vortex particles carrying orbital angular momentum (OAM) 
with respect to their average propagation direction \cite{Allen, Mono, Bliokh_07, Bliokh_11, Uchida, Verbeeck, McMorran}, 
the so-called Airy beams \cite{Airy, Airy_beam, Airy_Exp, Airy_El_Exp}, as well as their generalizations \cite{AB_2010, Zhao, Mathieu}. 
These novel states were experimentally realized for photons, for electrons with the energy of $200-300$ keV and, more recently, for cold neutrons \cite{neu}. 
They have already found numerous applications -- see, for example, Refs.\cite{Mono, Serbo, I_PRD, I_Phase, Ivanov_PRA_2012, Angstrom}. 
For a vortex electron, for instance, the phase looks like $\varphi = \ell \phi$ with $\phi$ being an azimuthal angle and $\ell \equiv \ell_z$ the OAM. 
That is why change of the phase's sign can be achieved by flipping the latter, $\ell \rightarrow -\ell$. 
 

A key quantity of interest in this study is a scattering asymmetry, which comes into play because the cross section is neither even nor odd in ${\bm b}_{\varphi}$. 
It is only moderately attenuated, $\mathcal A \propto \lambda_c/\sigma_b$, in both the methods we discuss and for elastic scattering it signifies a lack of an up-down symmetry 
in the angular distributions of scattered particles. This asymmetry is a Lorentz scalar and it can reach the values higher than $10^{-4} - 10^{-3}$ 
for well-focused electrons with the intermediate energies. Thus, one can in principle answer the question of how a Coulomb- or hadronic phase $\zeta_{fi}$
changes with the scattering angle or with $t$ by measuring the angular distributions of final particles. The system of units $\hbar = c= 1$ is used.

\section{The scattering asymmetry}\label{Asymm}

Let us consider a $2\rightarrow N_f$ head-on collision of two beams with the mean momenta $\langle{\bm p}\rangle_1, \langle{\bm p}\rangle_2$, 
their uncertainties $\sigma_1, \sigma_2$, the spatial widths $\sigma_{b,1}, \sigma_{b,2}$, with overall phases of the wave functions (in momentum representation) $\varphi_1, \varphi_2$, 
and let the centers of the beams be separated by an impact-parameter ${\bm b}$.

A quantitative measure for contribution of the phase $\zeta_{fi}$ to the number of events $dN$ is the following asymmetry
\begin{eqnarray}
& \displaystyle
\mathcal A = \frac{dN [{\bm b}_{\varphi}] - dN[-{\bm b}_{\varphi}]}{dN [{\bm b}_{\varphi}] + dN[-{\bm b}_{\varphi}]} = \frac{d\sigma [{\bm b}_{\varphi}] - d\sigma[-{\bm b}_{\varphi}]}{d\sigma [{\bm b}_{\varphi}] + d\sigma[-{\bm b}_{\varphi}]},
\label{Eq01}
\end{eqnarray}
which has a simple analytical form in the paraxial regime.
Indeed, in the chosen kinematics the asymmetry can depend only on the following vectors:
$\Delta {\bm u} = {\bm u}_1 - {\bm u}_2$, ${\bm b}_{\varphi}$, $(\frac{\partial}{\partial{\bm p}_1} - \frac{\partial}{\partial{\bm p}_2})\zeta_{fi}$,
and it must be a linear function of the two latter ones. The only true scalar that satisfies these criteria is
\begin{eqnarray}
\displaystyle
\mathcal A = \frac{2 \Sigma_1^2 \Sigma_2^2}{\Sigma_1^2 + \Sigma_2^2}\, \left [\frac{\Delta {\bm u}}{|\Delta {\bm u}|} \times \left [\frac{\Delta {\bm u}}{|\Delta {\bm u}|} \times\langle {\bm b}_{\varphi} \rangle\right ]\right ] 
 \cr 
\displaystyle \cdot \left (\frac{\partial \zeta_{fi}}{\partial \langle{\bm p}\rangle_{2}} 
- \frac{\partial \zeta_{fi}}{\partial \langle{\bm p}\rangle_{1}}\right ) + \mathcal O\left(\left (\frac{2 \Sigma_1^2 \Sigma_2^2}{\Sigma_1^2 + \Sigma_2^2}\right )^2\right)
,
\label{Eq1}
\end{eqnarray}
where ${\bm u}_{1,2} = \langle {\bm p}\rangle_{1,2}/\varepsilon_{1,2} (\langle{\bm p}\rangle_{1,2}), \varepsilon ({\bm p}) = \sqrt{{\bm p}^2 + m^2}$, and the vector $\langle {\bm b}_{\varphi} \rangle = {\bm b} - \langle \partial \varphi_1/\partial {\bm p}_1 \rangle + \langle \partial \varphi_2/\partial {\bm p}_2\rangle$ 
is averaged with the Gaussian distributions, $(\sqrt{\pi}\sigma_{1,2})^{-3}\,\exp\{-({\bm p}_{1,2} - \langle {\bm p}\rangle_{1,2})^2/\sigma_{1,2}^2\}$.
This asymmetry is invariant under Lorentz boosts along the collision axis and it vanishes in the plane-wave limit when either $\sigma_{1,2} \rightarrow 0$ or $\sigma_{b,1,2} \rightarrow \infty$.
Of course the main formula of this study, Eq.(\ref{Eq1}), can also be rigorously derived from a general expression for the scattering events: see the Appendix.

A small parameter of this series,
\begin{eqnarray}
\displaystyle
\frac{2 \Sigma_1^2 \Sigma_2^2}{\Sigma_1^2 + \Sigma_2^2} \equiv \mathcal O(\sigma_b^{-2}),\, \Sigma_{1,2}^2 = \frac{\sigma_{1,2}^2}{1 + \sigma_{1,2}^2 \sigma_{b,1,2}^2} \equiv \mathcal O(\sigma_{b,1,2}^{-2}),
\label{alpha}
\end{eqnarray}
appears thanks to finite overlap of the incoming packets. It is approximately $2/(\sigma_{b,1}^2 + \sigma_{b,2}^2)$ for wide beams 
with $\sigma_{b,1,2} \gg 1/\sigma_{1,2}$ and $\Sigma_{1,2} \approx 1/\sigma_{b,1,2}$.
This approximation is realized in the majority of practical cases. Say, for the LHC proton beam with $\sigma_b \sim 10 \mu$m 
and the monochromaticity of $\sigma/\langle p \rangle \lesssim 1\%$ \cite{LHC}, we have $\sigma \sigma_b > 10^{11}$. 

On the other hand, for non-relativistic beams the situation with $\sigma\sigma_b \gtrsim 1$ is also possible (see below), 
and in this case again $\Sigma_{1,2} \sim 1/\sigma_{b,1,2}$. Thus $\Sigma_{1,2}$ represent the momentum widths of the particle beams.

The averaging of ${\bm b}_{\varphi}$ has appeared because some phases $\varphi_{1,2}$ may not be analytical in the entire ${\bm p}$-domain, 
but contain a finite number of removable singularities. Say, for vortex beams with $\varphi = \ell \phi$ the derivative 
\begin{eqnarray}
& \displaystyle
\frac{\partial \varphi}{\partial {\bm p}} = \ell\, \frac{\hat{{\bm z}} \times {\bm p}}{{\bm p}_{\perp}^2}
\label{der}
\end{eqnarray}
is not analytical for a vanishing transverse momentum. This singularity is removable and the mean value of this,
\begin{eqnarray}
& \displaystyle
\Big \langle\frac{\hat{{\bm z}} \times {\bm p}}{{\bm p}_{\perp}^2} \Big \rangle = \frac{\hat{{\bm z}} \times \langle{\bm p}\rangle}{\langle {\bm p}_{\perp}\rangle^2}\left (1 - e^{-\langle {\bm p}_{\perp} \rangle^2/\sigma^2}\right ) 
,
\label{mean_value}
\end{eqnarray}
simply vanishes when $\langle{\bm p}_{\perp} \rangle \rightarrow 0$. 
Note that the number of events itself is suppressed as $\exp\{-\ell^2 \Sigma_{1,2}^2/(2{\bm p}_{\perp,1,2}^2)\}$ when ${\bm p}_{\perp,1,2}^2 \ll \Sigma_{1,2}^2$.

The asymmetry (\ref{Eq1}) depends on the final particles' momenta and in order to measure it, one should compare outcomes of the two experiments with a flipped sign of $\langle {\bm b}_{\varphi} \rangle$. 
This could be realized
\begin{itemize}
\item 
Either by swapping the two incoming beams with no phases whatsoever, that is, by ${\bm b} \rightarrow - {\bm b}$ 
(see Fig.\ref{Fig}),
\item
Or by changing the signs of the phases, $\varphi_{1,2} \rightarrow - \varphi_{1,2}$ (say, by $\ell \rightarrow - \ell$), with zero impact-parameter.
\end{itemize}
In what follows we shall discuss these means in detail. 
Note that Eq.(\ref{Eq1}) was obtained in the lowest order of perturbation theory with $\Sigma_{1,2} \ll |\langle {\bm p}\rangle_{1,2}|$; 
that is why all corrections due to the phases are supposed to be small anyway, 
that is, $|\mathcal A| \ll 1$ or, at the best, $|\mathcal A| \lesssim 1$. Otherwise this expression is inapplicable.

For a $2\rightarrow 2$ head-on elastic collision in the centre-of-mass frame of particles with the same masses and momentum widths $\Sigma \approx 1/\sigma_b$
(say, $pp \rightarrow p^{\prime}p^{\prime}$ or $ee \rightarrow e^{\prime}e^{\prime}$), we find from Eq.(\ref{Eq1}) 
\begin{eqnarray}
& \displaystyle
\mathcal A \approx - \frac{4}{\sigma_b^2}\, {\bm p}_3\, \frac{{\bm u} \times [{\bm u} \times \langle {\bm b}_{\varphi} \rangle]}{{\bm u}^2}\, \frac{\partial\, \zeta_{fi}}{\partial t},
\label{Eq2}
\end{eqnarray}
where we have put $\langle{\bm p}\rangle_{1} \equiv {\bm p} = {\bm u} \varepsilon = - \langle{\bm p}\rangle_{2}$. 
From Eq.(\ref{Eq2}) we infer that for the strictly forward scattering, ${\bm p}_3 \rightarrow {\bm p}$, the asymmetry vanishes.


\section{$1^{\text{st}}$ scenario: off-center collision of Gaussian beams}

In a first scenario with the two (in fact, not necessarily Gaussian) beams with no phases, 
one can put ${\bm u}_{\perp} = 0$ and $\langle {\bm b}_{\varphi} \rangle = {\bm b} =\{b, 0, 0\}$, 
where $b \sim \sigma_b$. In this case we find from Eq.(\ref{Eq2}):
\begin{eqnarray}
& \displaystyle
\mathcal A \approx 4\,\frac{p_3}{\sigma_b}\,\sin \theta_{sc}\cos\phi_{sc}\, \frac{\partial\, \zeta_{fi}}{\partial t}.
\label{Eq6}
\end{eqnarray}
This asymmetry is only linearly attenuated by $\sigma_b$ and it has a simple $\sin \theta_{sc}\cos\phi_{sc}$ dependence on the scattering angles $\theta_{sc}, \phi_{sc}$.
Any deviation of the measured asymmetry from this dependence would be an evidence of a non-trivial phase $\zeta_{fi} (s, t)$.

A numerical estimate of the asymmetry can be obtained for elastic scattering in the relativistic case 
with $p_3 \approx p,\,\, t \approx - p^2 \theta_{sc}^2,\, \theta_{sc} \ll 1,\, \gamma = \varepsilon/m \gg 1$. 
Assuming that the phase is a fast function of the scattering angle $\theta_{sc}$, but a slow one of $p$, we get
\begin{equation}
\displaystyle
\mathcal A \approx - 2\,\frac{1}{p \sigma_b}\cos\phi_{sc}\, \frac{\partial\, \zeta_{fi}}{\partial \theta_{sc}} \approx - 2\,\frac{\lambda_c}{\sigma_b}\cos\phi_{sc}\, \frac{1}{\gamma}\frac{\partial\, \zeta_{fi}}{\partial \theta_{sc}} \label{Eq6a}
\end{equation}
where $\lambda_c/\sigma_b$ and $\gamma^{-1}\,\partial \zeta_{fi}/\partial \theta_{sc}$ are Lorentz invariant separately.
As we have seen, it is the factor 
\begin{eqnarray}
& \displaystyle
\frac{\lambda_c}{\sigma_b} \ll 1
\label{Eq6ab}
\end{eqnarray}
that determines sensitivity to the phase and we wish to increase it. 
For protons, it is of the order of $10^{-10}$ for moderately relativistic beams focused to a spot of $\sim 1 \mu$m 
and it is $\sim 10^{-8}$ for protons with $p \approx 2$ MeV and focused to $\sigma_b \gtrsim 10$ nm \cite{p}. 
The estimate (\ref{Eq6a}), however, is inapplicable for such a non-relativistic case.

Conversely, in collision of electrons the ratio $\lambda_c/\sigma_b$ becomes bigger than $10^{-3}$ for $300$-keV beams focused 
in a spot of the order of $1 \text{\AA}$ \cite{Angstrom} (regardless of the OAM), even though the estimate (\ref{Eq6a}) 
can be used only for qualitative analysis for such intermediate energies. 
For the Coulomb phase on a one-loop level \cite{Y}
\begin{eqnarray}
& \displaystyle
\frac{1}{\gamma}\frac{\partial\, \zeta_{fi}}{\partial \theta_{sc}} \sim \frac{\alpha_{em}}{\gamma \theta_{sc}}
\label{zetatheta}
\end{eqnarray}
with $\alpha_{em} \approx 1/137$ and hence
\begin{eqnarray}
& \displaystyle
\mathcal A = \mathcal O \left (\frac{\lambda_c}{\sigma_b}\frac{\alpha_{em}}{\gamma\theta_{sc}}\right ).
\label{Eq13aaa}
\end{eqnarray}
This estimate is in accordance with that of the recent paper \cite{I_C} where what we call the $2^{\text{nd}}$ scenario is studied.
In the current scheme, we bring two sub-nm-sized electron beams into collision (note that in this case $1/\sigma \sim \sigma_b$), slightly off-center, 
and that is why one ought to be able to control their relative position with the accuracy better than $0.5 \text{\AA}$.
Then angular distributions of the scattered electrons are measured and compared in the upper- and in the lower semi-spaces.
Their difference reveals itself in the asymmetry and its conservative estimate for the scattering angles of $\theta_{sc} \sim 10^{-2}-10^{-1}$ is
\begin{eqnarray}
& \displaystyle
\mathcal |A| \sim 10^{-4} - 10^{-3},
\label{Aestim}
\end{eqnarray}
which is in principle measurable with high statistics. One could further increase it by performing measurements at yet smaller scattering angles 
or by making the impact parameter very large, $b \gg \sigma_b$.
In the latter case, however, the price is a drop in the number of events.

Returning to scattering of protons, little can be said independently of a model, unfortunately, about the factor in the left-hand-side of (\ref{zetatheta}). 
The TOTEM collaboration is able to perform measurements at the scattering angles smaller than $10^{-4}$ at $\sqrt{s} = 7$ TeV \cite{TOTEM}, which yields $\gamma \theta_{sc} \sim 0.1 - 1$, 
and the hadronic (or relative) phase $\zeta_{fi}$ itself, unlike the Coulomb one, is not attenuated by a small parameter $\alpha_{em} \rightarrow \alpha_s$, 
as scattering within a diffraction cone is not described by perturbation theory. This can, at least partly, compensate 
the lower value of $\lambda_c/\sigma_b$ and lead to a detectable effect for the hadronic phase. 
Anyway one should strive to make the beam's width $\sigma_b$ as small as possible.

Note that this asymmetry is a purely quantum effect that vanishes in the plane-wave limit 
and might seem to be counter-intuitive from a classical perspective.
Indeed, for a pair of azimuthally symmetric wave packets their substitution clearly does not alter the (classical) cross section. 
It is violated when either the packets are not-azimuthally symmetric 
(the $2^{\text{nd}}$ scenario) or the particles themselves have some inner structure (atoms, ions, hadrons). 
It is the latter case in which the phase $\zeta_{fi}$ comes into play.

\section{$2^{\text{nd}}$ scenario: colliding beams with phases} 


Within the second scenario, we start with a head-on collision of two vortex beams with ${\bm b} = 0$, 
the phases $\varphi_{1,2} = \ell_{1,2}\,\phi_{1,2}$, $\ell \equiv \ell_z$, 
and the opposite signs of their orbital helicities \cite{I_PRD}. 
The spatial distribution of such beams is no longer Gaussian but is a doughnut-shaped one with a minimum on the collision axis 
(see details in \cite{Uchida, Verbeeck, McMorran}).
Still working in the frame with $\langle{\bm p}\rangle_{1} \equiv {\bm p} = - \langle{\bm p}\rangle_{2}$, 
we find with the help of Eq.(\ref{mean_value}):
\begin{eqnarray}
& \displaystyle
\langle {\bm b}_{\varphi} \rangle = - (\ell_1 + \ell_2)\, \frac{\hat{\bm z} \times \langle{\bm p}\rangle}{\langle{\bm p}\rangle_{\perp}^2}\, (1 - e^{-\langle{\bm p}\rangle_{\perp}^2/\sigma^2}).
\label{Eq7}
\end{eqnarray}
This vector vanishes, together with the asymmetry, when either the total OAM of the system is zero, $\ell_1 + \ell_2 = 0$, 
or azimuthally symmetric beams with ${\bm u}_{\perp} = 0$ collide (say, the so-called pure Bessel beams, which are the simplest models of vortex states with an azimuthally symmetric profile \cite{Bliokh_11}). 
This takes place because in order to have a non-vanishing $\mathcal A$ azimuthal symmetry of the problem must be broken already in the initial state 
(exactly as in the previous example).

With vanishing impact-parameter, violation of the azimuthal symmetry can be achieved by shifting the phase vortex off a symmetry axis of the beam. 
When dealing with the holograms (as in Refs.\cite{Verbeeck, McMorran}), a shift of a fork dislocation off the beam center 
provides a (small) azimuthal asymmetry or, in other words, a non-vanishing transverse momentum (see details, for example, in \cite{Mair_2001}).
Such a shift is to be small, $\delta\rho \lesssim \sigma_b$, $\delta p_{\perp} \gtrsim 1/\sigma_b,\, \delta\theta \sim 1/(p\sigma_b)$
and it is made to opposite directions for both beams. To put it simply, 
in this scenario a non-vanishing transverse momentum plays the same role as does a finite impact parameter in the first one.

We arrive at the following estimate from Eq.(\ref{Eq2}): 
\begin{eqnarray}
& \displaystyle
& \displaystyle
\mathcal A \approx -4(\ell_1 + \ell_2) \frac{p_3}{\sigma_b^2 \sigma} \sin \theta_3 \sin (\phi_3 - \phi)\,
\frac{\partial\, \zeta_{fi}}{\partial t} 
\label{Eq8}
\end{eqnarray}
The major difference between this expression and Eq.(\ref{Eq6}) is appearance of the factor $\ell_1 + \ell_2$, which can be very large.
It might seem therefore that the second scenario with $\ell_{1,2} \gg 1$ provides a much higher value of the asymmetry.
However the price for this increase is again the drop in statistics due to the factor $\exp\{-\ell_{1,2}^2/(2\sigma_{b,1,2}^2{\bm p}_{1,2, \perp}^2)\}$
in Eq.(\ref{Eq6Alt}) in the Appendix. 

As the production of twisted hadrons with azimuthally non-symmetric profiles seems to be more technologically challenging than it is for electrons,
we turn to elastic scattering of the latter. By analogy with Eq.(\ref{Eq6a}), the factor  
\begin{eqnarray}
& \displaystyle
\frac{\ell_1 + \ell_2}{p \sigma_b^2\sigma}
\label{Eq8a}
\end{eqnarray}
determines the sensitivity to the asymmetry in the relativistic case. The maximum value of the OAM for which the number of events is not suppressed is 
$$
\ell_{\text{max}} \sim p_{\perp} \sigma_b \sim \sigma \sigma_b,
$$
and that is why 
\begin{eqnarray}
& \displaystyle
\mathcal A = \mathcal O \left (\lambda_c/\sigma_b\right),
\label{Eq8b}
\end{eqnarray}
exactly as in the previous scenario. Taking again $300$-keV twisted electrons focused to $\sigma_b \sim 1 \text{\AA}$,
with the monochromaticity of $\sigma/p \lesssim 1\%$, and $\sigma \sigma_b \sim 1$, we arrive at the same dependence (\ref{Eq8b}) 
when $\ell_{1,2} = \ell_{\text{max}} = 1$. 
In order to measure the asymmetry, the angular distributions of scattered electrons are to be compared
in the two experiments with $\ell_{1,2} = 1$ and $\ell_{1,2} = -1$, respectively. As before, one can alternatively carry out only one experiment with $\ell_{1,2} = 1$ 
when comparing angular distributions in the upper- and in the lower semi-spaces.
The numerical estimate (\ref{Aestim}) stays valid. Since for such a study we need vortex electrons with azimuthally asymmetric profiles, we would also like to find such states 
for which the requirement of a non-vanishing transverse momentum can be relaxed.



As can be readily seen, it is the case for Airy beams as their azimuthal distribution itself is not symmetric. 
For collision of two such states with ${\bm u}_{\perp} = 0$, the phase $\varphi = (\xi_x^3 p_x^3 + \xi_y^3 p_y^3)/3$, 
and the opposite signs of their parameters ${\bm \xi}_1 = -{\bm \xi}_2 \equiv {\bm \xi} = \{\xi, 0, 0\}$,
we find:
\begin{eqnarray}
&\displaystyle
\langle {\bm b}_{\varphi} \rangle = - \sigma^2 \{\xi^3, 0, 0\},\
\cr 
&\displaystyle \mathcal A \approx -4\, \frac{\sigma^2}{\sigma_b^2}\, \xi^3 p_3 \sin \theta_{sc}\cos \phi_{sc}\, \frac{\partial\, \zeta_{fi} }{\partial t},
\label{Eq10}
\end{eqnarray}
where we have used $\langle p_x^2 \rangle = \langle p_x \rangle^2 + \sigma^2/2$. The typical values of $\xi$ follow from the factor $\exp\{-\Sigma^2(\sigma^2\xi^3/2)^2/2\}$ 
in the probability formula. That is why
\begin{eqnarray}
& \displaystyle
\xi_{\text{max}} \sim \sigma_b\quad \text{when}\quad \sigma\sigma_b \sim 1,\cr
& \displaystyle \text{or}\quad \xi_{\text{max}} \sim \frac{\sigma_b}{(\sigma \sigma_b)^{2/3}} \ll \sigma_b\quad \text{when}\quad \sigma\sigma_b \gg 1.
\label{Eq10a}
\end{eqnarray}
In both cases, this yields the same $p_3/\sigma_b$ factor in the asymmetry as in Eq.(\ref{Eq6}) and $\lambda_c/\sigma_b$ for the relativistic energies.
Therefore the use of Airy beams leads to the very same predictions for the asymmetry as in the previous examples.

Moreover, one could think of such a phase $\varphi ({\bm p})$ that maximizes the asymmetry. Within the paraxial regime, however,
the phases are limited by the following inequality:
\begin{eqnarray}
& \displaystyle
\left \langle \left |\frac{\partial \varphi}{\partial {\bm p}}\right | \right\rangle \lesssim \sigma_b,
\label{Eq10aa}
\end{eqnarray}
which is simply analogous to $b \lesssim \sigma_b$. 
That is why the asymmetry stays $\mathcal O (\lambda_c/\sigma_b)$ for all other types of the non-plane-wave states as well.



The idea of using vortex states for probing the amplitude's phase was put forward by Ivanov \cite{I_Phase}. 
By analogy with his work, let us consider now scattering of a light particle by a heavy one (say, $e p \rightarrow X, \gamma p\rightarrow X$) with $\sigma_1/\sigma_2 \ll 1$.
Working in the frame in which the longitudinal momentum of the heavy particle is zero, we assume the light one to be in the pure Bessel state with ${\bm u}_{\perp,1} = 0$.
We obtain that the asymmetry,
\begin{eqnarray}
& \displaystyle
\mathcal A 
\propto \ell_2 \sigma_1\,\frac{\sigma_1}{\sigma_2},
\label{Eq13}
\end{eqnarray}
does not depend on the OAM $\ell_1$ of the light particle and, compared to Eq.(\ref{Eq8}), has an additional small factor $\sigma_1/\sigma_2$, 
which is less than $10^{-3}$ for the available beams. 
This factor also appears for the Airy beams when ${\bm p}_1 \ne -{\bm p}_2$ but $\sigma_1 \ll \sigma_2$. 
That is why the higher values of the asymmetry favor the case with $\sigma_1 \sim \sigma_2$, in accordance with the Ref.\cite{I_Phase}.

The difference between the two methods described above can be elucidated by comparing two ways of colliding two rubber balls. 
If the balls are pumped up well, they are azimuthally symmetric and in order to violate this symmetry in scattering we need to collide them slightly off-center.
Conversely, when the balls are deflated they are most likely no longer azimuthally symmetric and that is why they can collide even at a zero impact parameter.
One simply needs to imagine a wave packet with a non-trivial wave front instead of such a deflated ball.


\section{Summary}

As we have demonstrated, the cross section in a general scattering process 
becomes sensitive to the overall phase of the scattering amplitude
in a more realistic model with incoming particles described as wave packets. 
This phase reveals itself in the up-down angular asymmetry when either impact-parameter between the two beams 
is non-vanishing or the beams have non-trivial wave functions, that is, carry phases.
In both these scenarios, violation of the azimuthal symmetry of the problem,
which does not take place in the plane-wave approximation, 
yields a non-vanishing contribution of the phase $\zeta_{fi}$.

The asymmetry is only linearly attenuated by a small parameter $\lambda_c/\sigma_b$, regardless of the scenario. 
Its numerical estimate for the Coulomb phase is higher than $10^{-4}-10^{-3}$ for well-focused electrons with the intermediate energies. 
Corresponding experiments, albeit being challenging, can be performed at modern electron microscopes, 
both with the Gaussian beams and with the vortex- and/or Airy ones if they are focused to a spot of the order of or less than $1 \text{\AA}$ in diameter.

Predictions for the hadronic (or relative) phase are less certain and inevitably model-dependent.
Whereas the parameter $\lambda_c/\sigma_b$ does not exceed $10^{-10}$ even for the moderately relativistic proton beams, effects of the hadronic phase per se 
must be much stronger at small scattering angles (within the diffraction cone) than it is for the Coulomb phase, 
as the perturbation theory does not work there.
This can improve the chances for detecting the asymmetry.
As the twisted- or Airy protons have not been created yet, the corresponding experiments can be carried out within the first scenario.
On the other hand, generation of the fast but non-relativistic protons with the non-Gaussian wave functions (say, of the Airy ones) 
could facilitate the realization of these objectives.






\acknowledgments

I am grateful to E.\,Akhmedov, I.\,Ginzburg, I.\,Ivanov, P.\,Kazinski, G.\,Kotkin, V.\,Serbo, O.\,Skoromnik, A.\,Zhevlakov and, especially, to A.\,Di Piazza 
for many fruitful discussions and criticism. I also would like to thank C.\,H.\,Keitel, A.\,Di Piazza and S.\,Babacan for their hospitality 
during my stay at the Max-Planck-Institute for Nuclear Physics in Heidelberg. 
This work is supported by the Alexander von Humboldt Foundation (Germany) and by the Competitiveness Improvement Program of the Tomsk State University.

\section{Appendix: derivation of the asymmetry formula}

For a $2\rightarrow N_f$ scattering, the probability in a general non-plane-wave case according to Kotkin \textit{et al.}\cite{Impact} represents a functional 
of the (generalized) cross section $d \sigma ({\bm k}, {\bm p}_{1,2})$ and of a function that we shall denote $\mathcal L^{(2)} ({\bm k}, {\bm p}_{1,2})$ and call the particle correlator:
\begin{eqnarray}
& \displaystyle
dW = |S_{fi}|^2\, \prod\limits_{f=3}^{N_f+2} V \frac{d^3 p_f}{(2\pi)^3} =  \int \frac{d^3 p_1}{(2\pi)^3}\frac{d^3 p_2}{(2\pi)^3}\frac{d^3 k}{(2\pi)^3}
\cr & \displaystyle
\times d \sigma ({\bm k}, {\bm p}_{1,2})\, \mathcal L^{(2)} ({\bm k}, {\bm p}_{1,2}),\,\, d \sigma ({\bm k}, {\bm p}_{1,2}) = \cr & \displaystyle
= (2\pi)^4\, \delta \Big (\varepsilon_1 ({\bm p}_1 + {\bm k}/2) + \varepsilon_2 ({\bm p}_2 - {\bm k}/2) - \varepsilon_f \Big ) 
\cr & \displaystyle \times \delta^{(3)} ({\bm p}_1 + {\bm p}_2 - {\bm p}_f) 
\cr & \displaystyle 
\times T_{fi} ({\bm p}_1 + {\bm k}/2, {\bm p}_2 - {\bm k}/2) {T_{fi}^*} ({\bm p}_1 - {\bm k}/2, {\bm p}_2 + {\bm k}/2)
\cr & \displaystyle \times\frac{1}{\upsilon ({\bm p}_1, {\bm p}_2)}\prod\limits_{f=3}^{N_f+2} \frac{d^3 p_f}{(2\pi)^3},
\cr & \displaystyle 
\mathcal L^{(2)} ({\bm k}, {\bm p}_{1,2}) = \upsilon ({\bm p}_1, {\bm p}_2) \int\, dt\, d^3 r\,d^3 R\, e^{i{\bm k}{\bm R}}\, \cr & \displaystyle
\times n_1 ({\bm r}, {\bm p}_1, t) n_2 ({\bm r} + {\bm R}, {\bm p}_2, t),
\label{M}
\end{eqnarray}
where 
\begin{eqnarray}
& \displaystyle 
T_{fi} = \frac{M_{fi}}{\sqrt{2 \varepsilon_1 2 \varepsilon_2 \prod\limits_{f=3}^{N_f+2}2\varepsilon_f}},\ 
\cr & \displaystyle
\upsilon = \frac{\sqrt {(p_1 p_2)^2 - m_1^2 m_2^2}}{\varepsilon_1 ({\bm p}_1) \varepsilon_2 ({\bm p}_2)} = \sqrt{({\bm u}_1 - {\bm u}_2)^2 - [{\bm u}_1 \times {\bm u}_2]^2},
\cr & \displaystyle 
\varepsilon_f = \sum\limits_{i=3}^{N_f+2}\varepsilon_i ({\bm p}_i),\,\, \varepsilon ({\bm p}) = \sqrt{{\bm p}^2 + m^2},
\cr & \displaystyle
{\bm p}_f = \sum\limits_{i=3}^{N_f+2}{\bm p}_i,\,\, {\bm u}_{1,2} = {\bm p}_{1,2}/\varepsilon_{1,2} ({\bm p}_{1,2}),
\label{Ma}
\end{eqnarray}
and $n ({\bm r}, {\bm p}, t)$ is a particle's Wigner function with the normalization $ \int \frac{d^3p}{(2\pi)^3}\, d^3 r\, n ({\bm r}, {\bm p}, t) = 1$.

One can define an effective cross section by dividing the probability by a luminosity $L$,
\begin{eqnarray}
& \displaystyle d\sigma = \frac{dW}{L},\ L = \int \frac{d^3 p_1}{(2\pi)^3}\frac{d^3 p_2}{(2\pi)^3}\frac{d^3 k}{(2\pi)^3}\,\, \mathcal L^{(2)} ({\bm k}, {\bm p}_{1,2}) = 
\cr & \displaystyle
= \int \frac{d^3 p_1}{(2\pi)^3}\frac{d^3 p_2}{(2\pi)^3}\,\, dt d^3 r\,\, \upsilon\, n_1 ({\bm r}, {\bm p}_1, t) n_2 ({\bm r}, {\bm p}_2, t).
\label{Cr}
\end{eqnarray}
It is this quantity that we are going to compare the plane-wave cross section with.

For a pure state, the Wigner function is
\begin{eqnarray}
\displaystyle
n ({\bm r}, {\bm p}, t) = \int \frac{d^3k}{(2\pi)^3}\, e^{i{\bm k}{\bm r}}\, \psi^* ({\bm p} - {\bm k}/2, t) \psi ({\bm p} + {\bm k}/2, t),
\label{WA}
\end{eqnarray}
where $\psi ({\bm p}, t) = \psi ({\bm p})\, \exp\{-i t\, \varepsilon ({\bm p})\}$.
As an example, we take a Gaussian wave packet with the phase $\varphi ({\bm p})$:
\begin{eqnarray}
\displaystyle
\psi ({\bm p}) = \pi^{3/4} \Big ({\frac{2}{\sigma}}\Big )^{3/2}\, \exp \Big \{ - i{\bm r}_{0}{\bm p} -\frac{({\bm p} - \langle{\bm p}\rangle)^2}{2\sigma^2} + 
\cr \displaystyle + i \varphi ({\bm p})\Big\}. 
\label{psivarphi}
\end{eqnarray}
where ${\bm r}_0$ denotes a center of the wave packet at $t=0$. 
Then the Wigner function is
\begin{eqnarray}
\displaystyle
n = \frac{1}{(\sqrt{\pi}\, \sigma)^3}\int d^3k \exp\Big\{-\frac{({\bm p} - \langle {\bm p}\rangle)^2}{\sigma^2} - \frac{{\bm k}^2}{(2\sigma)^2} + 
\cr \displaystyle 
+ i {\bm k} ({\bm r} - {\bm r}_0) 
-it \left (\varepsilon ({\bm p} + {\bm k}/2) - \varepsilon ({\bm p} - {\bm k}/2)\right ) +
\cr \displaystyle
+ i \Big (\varphi ({\bm p} + {\bm k}/2) - \varphi ({\bm p} - {\bm k}/2)\Big )\Big\}.
\label{Eq1Alt}
\end{eqnarray}

In a real experiment the beams of $N_b$ particles collide, and their widths, $\sigma_b$, may be many orders of magnitude higher than a coherence length of one wave packet $1/\sigma$. 
Taking as an example two Gaussian beams with $N_{b,1}, N_{b,2}$ particles and the corresponding distribution functions, 
$\pi^{-3/2}\sigma_b^{-3}\exp\{-({\bm r}_0 - {\bm r}_b)^2/\sigma_b^2\}$ with ${\bm r}_b$ pointing to the beam's center at $t=0$, 
we perform statistical averaging of the one-particle correlator from Eq.(\ref{M}). In doing so, we imply that the mean distance between particles in the beams, 
$\sim \sigma_b/N_b$, does not exceed the coherence length $1/\sigma$, i.e. $ N_b \gtrsim\, \text{or}\, \gg \sigma \sigma_b$.
The result of the averaging is
\begin{eqnarray}
\displaystyle
{\mathcal L}_b = N_{b,1} N_{b,2} \frac{(2\pi)^7 \upsilon}{(\pi\, \sigma_1 \sigma_2)^3}\, \delta \Big (\varepsilon_1 ({\bm p}_1 + {\bm k}/2) - 
\cr \displaystyle
- \varepsilon_1 ({\bm p}_1 - {\bm k}/2) + \varepsilon_2 ({\bm p}_2 - {\bm k}/2) - \varepsilon_2 ({\bm p}_2 + {\bm k}/2)\Big )
\cr \displaystyle
\times \exp\Big\{-\frac{({\bm p}_1 - \langle {\bm p}\rangle_1)^2}{\sigma_1^2} - \frac{({\bm p}_2 - \langle {\bm p}\rangle_2)^2}{\sigma_2^2} - 
\cr \displaystyle
- \left(\frac{{\bm k}}{2}\right)^2\left (\frac{1}{\Sigma_1^2} + \frac{1}{\Sigma_2^2}\right ) - i {\bm k} {\bm b} + 
\cr \displaystyle
+ i \Big (\varphi_1 ({\bm p}_1 + {\bm k}/2) - \varphi_1 ({\bm p}_1 - {\bm k}/2) + 
\cr \displaystyle
+ \varphi_2 ({\bm p}_2 - {\bm k}/2) - \varphi_2 ({\bm p}_2 + {\bm k}/2)\Big )\Big\}.
\label{Corrbeam}
\end{eqnarray}
where the relative impact parameter of the two beams, 
$$
{\bm b} = {\bm r}_{b,1} - {\bm r}_{b,2} = \{{\bm b}_{\perp}, 0\},
$$
is introduced and $\Sigma_{1,2}$ are from Eq.(\ref{alpha}).

As a result, we arrive at the formula (\ref{Eq3Alt}) for the number of scattering events.
\begin{widetext}
\begin{eqnarray}
& \displaystyle
dN = \prod\limits_{f=3}^{N_f+2}\frac{d^3p_f}{(2\pi)^3}\,N_{b,1} N_{b,2} \frac{(2\pi)^{11}}{(\pi\, \sigma_1 \sigma_2)^3}\, \int\frac{d^3p_1}{(2\pi)^3}\frac{d^3p_2}{(2\pi)^3}\frac{d^3 k}{(2\pi)^3}\,\,
\delta \left (\varepsilon_1 ({\bm p}_1 - {\bm k}/2) + \varepsilon_2 ({\bm p}_2 + {\bm k}/2) - \varepsilon_f\right )
\cr & \displaystyle
\times \delta \left (\varepsilon_1 ({\bm p}_1 + {\bm k}/2) + \varepsilon_2 ({\bm p}_2 - {\bm k}/2) - \varepsilon_f\right ) \delta ({\bm p}_1 + {\bm p}_2 - {\bm p}_f)\,
T_{fi} ({\bm p}_1 + {\bm k}/2, {\bm p}_2 - {\bm k}/2) T_{fi}^*({\bm p}_1 - {\bm k}/2, {\bm p}_2 + {\bm k}/2)
\cr & \displaystyle
\times \exp\Big\{-\frac{({\bm p}_1 - \langle {\bm p}\rangle_1)^2}{\sigma_1^2} - \frac{({\bm p}_2 - \langle {\bm p}\rangle_2)^2}{\sigma_2^2} - \left(\frac{{\bm k}}{2}\right)^2\left (\frac{1}{\Sigma_1^2} + \frac{1}{\Sigma_2^2}\right ) -
\cr & \displaystyle - i {\bm k} {\bm b} + i \left (\varphi_1 ({\bm p}_1 + {\bm k}/2) - \varphi_1 ({\bm p}_1 - {\bm k}/2) + \varphi_2 ({\bm p}_2 - {\bm k}/2) - \varphi_2 ({\bm p}_2 + {\bm k}/2)\right )\Big\}.
\label{Eq3Alt}
\end{eqnarray}
\end{widetext}
It can be further simplified in the paraxial regime with $\Sigma_{1,2}^2 \ll \langle {\bm p} \rangle_{1,2}^2$. 
To this end, we use integral representations for the energy delta-functions, $\delta (\varepsilon) \delta (\varepsilon^{\prime}) = \int \frac{d\tau}{2\pi} \frac{d\tau^{\prime}}{2\pi} \exp\{i\tau\varepsilon + i\tau^{\prime}\varepsilon^{\prime}\}$, expand all the functions under the integral into ${\bm k}$-series up to the 2nd order inclusive, 
thus keeping terms not higher than $O(\sigma_b^{-2})$, and then integrate over ${\bm k}$. 
In doing so, we imply that the amplitude $T_{fi}$ and the phases $\varphi_{1,2}$ are analytical- and slowly varying functions of the momenta
\footnote{A case when $\partial\varphi_{1,2}/\partial{\bm p}_{1,2}$ contains a removable singularity is discussed in the main text.} 
and use the following formulas:
\begin{eqnarray}
\displaystyle
\int d^3k \exp \left\{-k_i A_i -\frac{1}{2} B_{ij} k_i k_j\right\} \left\{1, k_m k_n\right\} 
= 
\cr \displaystyle
= \frac{(2\pi)^{3/2}}{\sqrt{\det B}}\, \exp \left\{\frac{1}{2} B^{-1}_{ij} A_i A_j\right\}
\Big\{1,\, B^{-1}_{m n} + 
\cr \displaystyle
+ B^{-1}_{mi}B^{-1}_{nj} A_iA_j\Big\},
\cr \displaystyle
T_{fi} ({\bm p}_1 + {\bm k}/2, {\bm p}_2 - {\bm k}/2) T_{fi}^*({\bm p}_1 - {\bm k}/2, {\bm p}_2 + {\bm k}/2) 
\approx 
\cr \displaystyle
\approx \Big (|T_{fi}|^2 + \frac{1}{4}\, k_i k_j\, C_{ij} + \mathcal O(k^4)\Big )\,
\cr \displaystyle
\times \exp\Big\{i{\bm k} \partial_{\Delta {\bm p}}\, \zeta_{fi} + \mathcal O(k^3)\Big\},
\cr \displaystyle
\partial_{\Delta {\bm p}} = \frac{\partial}{\partial {\bm p}_1} - \frac{\partial}{\partial {\bm p}_2},\,\, \Delta {\bm u} = {\bm u}_1 - {\bm u}_2,
\cr \displaystyle
C_{ij} = |T_{fi}| \partial_{\Delta {\bm p}_i}\partial_{\Delta {\bm p}_j}|T_{fi}| - (\partial_{\Delta {\bm p}_i}|T_{fi}|)(\partial_{\Delta {\bm p}_j}|T_{fi}|)
\cr \displaystyle
B_{ij} = \alpha\, \delta_{ij} + \beta\, u_{1,i}u_{1,j},\ 
\alpha = \frac{1}{2\Sigma_1^2} + \frac{1}{2\Sigma_2^2} - 
\cr \displaystyle
- \frac{i\tau}{4} \left(\frac{1}{\varepsilon_1} + \frac{1}{\varepsilon_2}\right),\,
\beta = \frac{i\tau}{4} \left(\frac{1}{\varepsilon_1} + \frac{1}{\varepsilon_2}\right),
\cr \displaystyle
\det B = \alpha^2 (\alpha + \beta {\bm u}_1^2),\, B^{-1}_{ij} = \frac{1}{\alpha} \Big (\delta_{ij} - 
\cr \displaystyle
- u_{1,i} u_{1,j} \frac{\beta}{\alpha + \beta {\bm u}_1^2}\Big ).
\label{Eq4bAlt}
\end{eqnarray}

The remaining integral over $\tau^{\prime}$ is easily evaluated, and the final result is:
\begin{eqnarray}
\displaystyle
dN = \prod\limits_{f=3}^{N_f+2}\frac{d^3p_f}{(2\pi)^3}\,N_{b,1} N_{b,2} \frac{(2\pi)^9}{(\pi\sigma_1 \sigma_2)^3} 
\cr \displaystyle
\times \int\frac{d^3p_1}{(2\pi)^3}\frac{d^3p_2}{(2\pi)^3}\frac{d\tau}{2\pi}\,
\frac{\delta ({\bm p}_1 + {\bm p}_2 - {\bm p}_f)}{\sqrt{\det B\, \Delta u B^{-1} \Delta u}}
\cr \displaystyle 
\times \Bigg (|T_{fi}|^2 + \frac{1}{4} C_{ij} \Big [B^{-1}_{ij} 
- B^{-1}_{im}B^{-1}_{jn} \Big(\tilde{b}{_{\varphi}}_m \tilde{b}{_{\varphi}}_n
- 
\cr \displaystyle
- 2 \Delta u_m \tilde{b}{_{\varphi}}_n\, \frac{\Delta u B^{-1}\tilde{b}_{\varphi}}
{\Delta u B^{-1} \Delta u} + \frac{\Delta u_m \Delta u_n}{\Delta u B^{-1} \Delta u} \Big (1 + 
\cr \displaystyle
+ \frac{(\Delta u B^{-1} \tilde{b}_{\varphi})^2}{\Delta u B^{-1} \Delta u}\Big )\Big )\Big ]\Bigg )
\exp\Big\{i\tau (\varepsilon_1 ({\bm p}_1) + \varepsilon_2 ({\bm p}_2) - \varepsilon_f) 
- 
\cr \displaystyle
- \frac{({\bm p}_1 - \langle {\bm p}\rangle_1)^2}{\sigma_1^2} - \frac{({\bm p}_2 - \langle {\bm p}\rangle_2)^2}{\sigma_2^2} 
+\frac{1}{2} \frac{(\Delta u B^{-1} \tilde{b}_{\varphi})^2}{\Delta u B^{-1} \Delta u} - 
\cr \displaystyle
- \frac{1}{2}\,\tilde{b}_{\varphi} B^{-1}\tilde{b}_{\varphi} \Big\},
\label{Eq6Alt}
\end{eqnarray}
where
\begin{eqnarray}
& \displaystyle
\cr & \displaystyle
\tilde{{\bm b}}_{\varphi}= {\bm b} - \frac{\partial \varphi_1}{\partial {\bm p}_1} + \frac{\partial \varphi_2}{\partial {\bm p}_2}  - \left(\frac{\partial}{\partial {\bm p}_1} - \frac{\partial}{\partial {\bm p}_2}\right) \zeta_{fi}
\label{Eqb}
\end{eqnarray}
is denoted.

The resultant integrand in (\ref{Eq6Alt}) can be expanded into $\Sigma_{1,2} (\approx 1/\sigma_{b,1,2})$-series when spreading of the beams during the collision is small (but non-vanishing!),
that is,
\begin{eqnarray}
& \displaystyle
t_{\text{col}} \ll t_{\text{diff}} \sim \frac{\sigma_b}{u_{\perp}} \sim \sigma_b^2 \varepsilon.
\label{t}
\end{eqnarray}
Indeed, one can neglect $\beta (\tau)$ compared to $1/(2\Sigma_1^2) + 1/(2\Sigma_2^2)$ in $|\alpha|$ in Eq.(\ref{Eq4bAlt})
exactly when $ |\tau| \ll 2 \sigma_b^2 \varepsilon$ for two identical beams with $\sigma_{b,1} = \sigma_{b,2} = \sigma_b \gg 1/\sigma_{1,2},\, \varepsilon_1 = \varepsilon_2 = \varepsilon$.
Following this procedure, we get series $dN = dN^{(pw)} + dN^{(1)} + ...$ where the first term, as can be readily checked, 
yields the conventional plane-wave result, $dN^{(pw)} = L^{(pw)} d\sigma^{(pw)}$ with $L^{(pw)} = 2\Sigma_1^2\Sigma_2^2 \upsilon/(2\pi |\Delta {\bm u}| (\Sigma_1^2 + \Sigma_2^2))$,
and it is the first correction, $dN^{(1)}$, that contains all dependence on the phases $\zeta_{fi}$ and $\varphi_{1,2}$ that does not take place in the plane-wave approximation.
A similar expansion can also be made for the luminosity, $L \approx L^{(pw)} + L^{(1)}$ and for the cross section, $d\sigma = dN/L \approx d\sigma^{(pw)} + d\sigma^{(1)}$. 
We shall present these rather cumbersome formulas together with a more detailed discussion elsewhere.
A small parameter of these series appears to be the one from Eq.(\ref{alpha}). 
Substituting this expansion into Eq.(3) in the paper, we finally arrive at the main formula for the asymmetry, Eq.(4).

\end{document}